# Predicting Readiness to Engage in Psychotherapy of People with Chronic Pain Based on their Pain-Related Narratives

Saar Draznin Shiran[1], Boris, Boltyansky[1], Alexandra Zhuravleva[1], Dmitry Scherbakov[1], Pavel Goldstein[1]

1. Integrative Pain Laboratory, School of Public Health, University of Haifa, Israel

## Abstract

**Background**: Chronic pain affects 20% of the worldwide population. The classical biomedical approach to pain led many to seek solely medical cure to their pain, leading to a major opioid crisis. A paradigm shift towards the biopsychosocial model recognizes chronic pain's subjective and complex nature. Despite psychosocial interventions addressing this complexity, a gap exists between their effectiveness and patients' readiness to engage in psychosocial interventions. This study aimed to assess individuals' perceptions and readiness for psychosocial interventions, hypothesizing that subjective pain-related narratives can predict readiness to engage in psychotherapy.

**Methods**: Using a quantitative cross-sectional design, we collected pain-related narratives from 24 chronic pain patients using a dedicated platform (Painstory.science), with a unique capability to record users' narratives. Narratives collected included open-ended questions regarding perception of pain source, pain interference, and factors affecting pain. The collected data was preprocessed with Natural Language Processing (NLP) approaches and a pretrained Large Language Model was used to extract meaningful patterns and features, predicting readiness to engage in psychotherapy.

**Results**: The study achieved impressive accuracy rates in predicting readiness to engage in psychotherapy using the narratives with a focus on three pain related domains. Perception of pain with accuracy of 95.65%, specificity of 0.8, sensitivity of 1, and a robust ROC curve for binary classification registering at 0.90. Factors influencing pain with accuracy of 83.33%, specificity of 0.6, sensitivity of 0.9, and a substantial corresponding ROC curve for binary of 0.75. Pain interference with low accuracy of 65.22%, specificity 0.2, sensitivity 0.8 while ROC curve registered at 0.49. Moreover, Correlation analyses revealed weak to modest positive associations between linguistic factors (sentence count) and readiness to engage in psychotherapy for pain perception narratives (r=0.54, p<0.01) and factors affecting pain (r=0.24, p<0.5).

**Conclusion**: In addressing the gap between psychosocial interventions and patient readiness, our study modestly contributes to understanding the prediction of engagement in psychotherapy. The findings suggest the potential of subjective pain narratives as indicators for involvement in psychosocial interventions, offering a promising avenue for enhancing patient-centered strategies in chronic pain management through the use of innovative methods such as Natural Language Processing (NLP) and Machine Learning (ML).

**Predicting Readiness to Engage in Psychotherapy of People with Chronic Pain Based on**

**their Pain-Related Narratives**

## 1. Introduction

Pain is an essential mechanism indicating that our body might be in danger, yet pain can become chronic and no longer a reliable indicator of direct tissue damage (Lumley et al., 2011). Today chronic pain affects the quality of life and well-being of millions around the world on many levels - physical (restrictions in mobility), emotional (anxiety and depression) and social (reduced quality of life). Moreover, chronic pain is the main cause of the current opioid dependency and overdose crisis (Dahlhamer et al., 2018; Shim et al., 2019; Torres et al., 2020), a problem leading researchers to try to understand the complexity of chronic pain and the factors that lead to chronic pain (Gatchel et al., 2018).

### 1.1 Models of Pain

The classical biomedical model of pain holds a simple logical view on pain and looks mainly at the pathophysiological and mechanical disease processes (Turk & Okifuji, 2002). In this view, chronic pain is perceived in a homogeneous way, lacking the importance of identifying different types of pain (Lumley & Schubiner, 2019), as well as the subjective factors of chronic pain patients. These include healthcare professionals' beliefs and attitudes, treatment orientations and expectations to maintain trust in the clinician– patient alliance (Ng et al., 2021; Turk, 1999). These factors can impact how patients present themselves and respond to the treatments offered to them. While it is important to address these factors, physicians in current health care systems are limited in time and training (Sajid, 2018) and often find dealing with the psychosocial issues patients come in with uncomfortable and burdensome (Tait et al., 2021), resulting in pain disbelief and sending the message that the patient's disease is 'all in their heads'(Saperia & Swartzman, 2012; Sarzi-Puttini et al., 2021). These challenges have a great impact on the way chronic pain sufferers feel and perceive their own pain, whereas lack of compassion may lead them to feel stigmatized, increased self-judgment and avoid seeking or adhering to pain management programs and psychosocial interventions (Nicola et al., 2021).

Today, pain takes a multidimensional view addressing the importance of the subjective experience of one's pain as well as non verbal behaviors of pain: "Pain is a distressing experience associated with actual or potential tissue damage with sensory, emotional, cognitive, and social components" (Williams & Craig, 2016). The biopsychosocial model of pain addresses these issues by viewing pain and illness in a more holistic manner, integrating the biomedical aspects together with social and psychological aspects that may impact or prolong chronic pain (Turk & Okifuji, 2002).

## 1.2    Recent Advances in Chronic Pain Research

Recent brain research reveals how chronic pain differs from acute pain in brain activation, showing that in chronic pain, areas in the brain that are associated with negative emotion and perception are activated; how this activation is influenced by the length of pain condition and how emotions and perception may play an important role in the transition from acute pain to chronic pain (Baliki et al., 2006; Hashmi et al., 2013). Moreover, digital health research has demonstrated the associations between negative and positive affect and pain levels (Goldstein et al., 2020). These developments reflected the need for a paradigm shift in chronic pain perception. As a result of this need, chronic pain recently received a coding system classifying chronic pain as a disease for the first time by The International Classification of Diseases, 11th Revision (ICD-11) (Smith et al., 2019).

## 1.3    Psychotherapeutic Approaches in Pain Management

In recent years numerous psychosocial interventions and therapies were developed or adapted for treating chronic pain and pain management based on the biopsychosocial model addressing the complexity and range of factors affecting chronic pain. Moreover, reviews show the importance of addressing emotional components in the treatment of chronic pain (Lumley & Schubiner, 2019). Several types of psychotherapies specialized in pain management aiming at the treatment of psychological components of pain (Csaszar et al., 2014). Treatments addressing these needs include classical CBT psychotherapy, Mind-Body Interventions (Keefe et al., 2013), Dance Movement Therapy (DMT) (Shim et al., 2019), Pain Reprocessing Therapy (PRT) (Ashar et al., 2022), Trauma Focused Psychotherapies and more (Lumley & Schubiner, 2019). Although the wide range of therapies, recent reviews show mixed results with decreased treatment efficacy, low engagement and high dropout rates demonstrating the diversity in treatment efficacy. Previous research has shown that patients themselves experience difficulties in accepting psychosocial explanations for their pain and tend to develop narrative strategies intended to reduce the chance of being classified as 'psychological' cases (May et al., 2000). Furthermore, patients decline to engage in psychosocial interventions possibly due to the stigma around mental health and the required more active approach of psychosocial interventions and chronic pain management (Driscoll et al., 2021).

These findings suggest that although the models of pain has developed and research has advanced, chronic pain sufferers themselves may need to make a shift in their pain mindset, accepting and acknowledging the complexity of chronic pain and the great impact of the subjective pain experience influencing one's readiness to engage in psychosocial interventions (Morley et al., 2013; Sharpe & Carson, 2001). There is therefore a need to assess chronic pain sufferers' perceptions regarding their condition and their pain experience as well as their readiness to engage in psychosocial interventions, as this may be crucial for the acceptance of psychosocial aspects of chronic pain and the engagement in psychosocial interventions.

## 1.4     Pain Mindset and Readiness to Engage in Psychosocial Interventions

Research shows that perception affects one's response on a variety of levels (Kassam et al., 2009) and that different stages of readiness to change influence behavioral changes (Katz et al., 2019). Pain appraisals are based on the meaning one gives to them and then acts upon these appraisals (Turk, 1999). Beliefs have an important role in shaping action and reaction to situations creating self-theories which will be either fixed or malleable. Malleable theories include openness to learning and confronting challenges (Dweck, 2008). Our mindset can affect how we understand and act in response to a situation. Holding enhancing type information may change cognitive and emotional response, thus affecting one's mindset (Crum et al., 2013, 2017). Educating patients regarding pain processing through pain education aims to change preexisting beliefs (Leake et al., 2021). Conceptual change theory challenges existing knowledge to help chronic pain patients, proposing that pain can be a protective body signal and not necessarily a direct marker of tissue damage (Fletcher et al., 2021; Moseley & Butler, 2015). Despite these efforts, chronic pain patients often continue to look for simple biomedical diagnosis for their pain and this is mainly caused by how they perceive their pain and its cause, thus greatly affecting their healing process (Bonfim et al., 2021). In addition, it is important to note the impact of clinicians' use of labels and terms on the perceptions of pain of their patients, and how this affects one's choice of treatment and openness to a biopsychosocial approach (O'Keeffe et al., 2022).

Assessing one's pain perceptions, and readiness to change requires understanding the contextual frame of the subjective pain experience. Patient's narratives may reveal important information about the type of pain and psychosocial features of chronic pain (Lumley & Schubiner, 2019), providing a rich and comprehensive form for expressing subjective experience, revealing highly sensitive personal context (Wideman et al., 2019). Moreover, recorded narratives may reveal nonverbal behavior, such as bodily gestures (Heath, 2002) and vocal indicators that patients employ when discussing issues concerning chronic pain (Torres et al., 2020). Recent research on psychosocial interventions emphasize the importance of assessing one's perceptions and readiness before offering them to engage in psychosocial interventions. Discovering patients characteristics and indicators of pain perception and readiness to engage in psychosocial interventions will enable optimizing personalized based treatment for chronic pain (Turk et al., 2011). Thus, collecting pain narratives of people with chronic pain will enable access to a large amount of data in an efficient way in order to assess subjects' perceptions and readiness to engage in psychosocial interventions. In this study our goal was to evaluate whether the linguistic content of pain narratives of people with chronic pain predict readiness to engage in psychosocial interventions, specifically psychotherapy, in order to personalize available pain interventions.

## 1.5    Importance of Research and Clinical Implications

The complex nature of chronic pain conditions leads to an understanding that the simple biomedical approach does not address psychosocial factors of chronic pain and is insufficient in the treatment of chronic pain. Although the biopsychosocial model and interventions based on this model aim to address and fix this deficiency, there is still a gap between the therapies and interventions offered today to people with chronic pain and one's readiness to accept this new psychosocial definition to their pain condition and engage in psychosocial interventions. This study tried to lead to a better conceptual understanding of components of chronic pain sufferers' mindset and perceptions and identify one's readiness to engage in psychotherapy for chronic pain. Better assessment may help health care professionals understand a patient's state and starting point for psychotherapy and thus better tailor personalized treatments to match the current needs of each individual. In the clinical field, taking in account the subjective experience through personal narratives in the process of diagnosis and tailoring treatments sends a new message to patients that their subjective experience is valid and important and thus reduces the existing stigma that chronic pain is not real and that the problem is "all in their heads".

## 2. Research Question

The present study examines whether pain narratives of people with chronic pain predict readiness to engage in psychotherapy?

Specifically:

A.   Whether narrative-based perception of pain source predicts readiness to engage in psychotherapy.

B. Whether narrative-based perception of pain interference predicts readiness to engage in psychotherapy.

C. Whether narrative-based factors increasing/decreasing pain predict readiness to engage in psychotherapy.

## 3. Hypotheses

We hypothesize pain narratives can predict readiness to engage in psychotherapy.

Specifically:

A.  Narrative-based perception of the pain source predicts readiness to engage in psychotherapy with an accuracy of 70%.

B.  Narrative-based perception of pain interference predicts readiness to engage in psychotherapy with an accuracy of 70%.

C. Narrative-based factors increasing/decreasing pain predict readiness to engage in psychotherapy with an accuracy of 70%.

# 4. Methods

## 4.1. Procedure

The study adopted a quantitative research approach with a cross-sectional design, employing a digital survey format on Painstory.science (see link), to systematically collect data and evaluate associations between pain narratives, psychotherapy readiness, and pain perception. We have developed this digital platform to allow us to collect audio-recorded pain narratives of people with chronic pain talking about their current symptoms and related emotions outside the clinical setting in a natural environment.

For participant recruitment, convenient sampling was employed, utilizing social media platforms. Participants engaged with the Painstory.Science platform via a provided link on their smartphones. During the survey, participants answered questions and recorded their pain narratives, dedicating around 2 minutes to each open-ended question. The entire survey concluded within 15-20 minutes. Prior to participation, all subjects signed a digital consent form, and for enhanced transparency, a concise video overview of the project was presented on the landing page. Confidentiality regulations were strictly followed, ensuring secure storage of all collected data. Through this platform, we aimed to capture authentic and contextually rich pain narratives, enabling a nuanced exploration of subjects' perceptions of their pain and readiness to engage in psychotherapy. The 'PainStory' platform collects information about patients' socio-demographic background, diagnosis, pain intensity and symptoms, pain perceptions, emotional state and readiness to engage in psychotherapy through questionnaires and audio recorded responses. The current study is focused on predicting readiness to engage in psychotherapy based on the recorded pain-related narratives collected.

## 4.2    Sample

The sample included adults (ages 18+), who met the general criteria of chronic pain: people suffering from pain for at least three months with pain levels above 3 on the Visual Analog Scale (VAS) (Boonstra et al., 2008). The dataset consisted of 24 respondents, most of the participants were females (75%) and mostly originated from the United States (57%), the majority of participants held either a bachelor's degree or a master's degree (67%) (see table 1).

| Individual-Level Variables | N | % | Mean (SD) |
|---|---|---|---|
| Age | 24 (Total) | | 48(16.6) |
| **Gender** | | | |
| Female | 18 | 75% | |
| Male | 6 | 25% | |
| **Education Level** | | | |
| Bachelor Degree | 8 | 33.33% | |
| Master's Degree | 8 | 33.33% | |
| School Level | 4 | 16.67% | |
| Professional Training | 3 | 12.5% | |
| Doctorate | 1 | 4.17% | |
| **Country Distribution(N=23)** | | | |
| United States (US) | 13 | 56.52% | |
| Israel | 5 | 21.74% | |
| Canada | 3 | 13.04% | |
| United Kingdom (UK) | 1 | 4.35% | |
| Pakistan | 1 | 4.35% | |

**Table1: Socio-Demographic Table Overall (N=24)**

## 4.3    Variables

For this study, we used open-ended questions and responses (also mentioned as "narratives") that were obtained from the speech recordings. Each of the open-ended questions was based on widely validated self-report questionnaire**s** and modified by a committee of four experts in the research field of chronic pain.

To assess the predictive validity of the proposed study, we populated the narratives into three groups as per hypothesis (A: *Narrative-Based Perception of pain source*, B: *Narrative-Based Pain interference*, C: *Narrative-Based factors affecting pain*), and analyzed those using the State-Of-The-Art Large Language Model to make the prediction on the readiness to engage psychotherapy in the future.

### 4.31 *Predictors*

**Narrative-Based Perception of Pain Source.** An open-end question: "What are the most important factors that you believe caused your pain in the beginning? How do you believe it became persistent?". For measuring perception of pain source, we chose to use a modified version of the ninth qualitative open-ended question from the widely validated self-report questionnaire, the Brief Illness Perception Questionnaire (B-IPQ) (Broadbent et al., 2006). The original format ("rank 3 most important factors that you believe caused your illness") was slightly modified in order to fit the invitation to respond in the form of a story. Subjects were given a window of two minutes to record their response.

**Narrative-Based Pain Interference.** An open-end question: "How does your pain affect your life - work, family, relationships?".  The question we chose is based on the PROMIS pain interference scale (Amtmann et al., 2010). Subjects were given a window of two minutes to record their response.

**Narrative-Based Factors Affecting Pain.** Two open-end questions included: 1. "What things that you do or experience in life make the pain worse? Describe physical and emotional triggers". 2. "What things that you do or experience in life make the pain better"? Subjects were given a window of two minutes to record their response.

### 4.32 *Outcome*

**Readiness to Engage in Psychotherapy.** For prediction if the user will be ready to engage in psychotherapy, the following question was used (self-reported): "Which of the following methods are you willing to try for your pain? Check all that apply". The current study focused only on psychotherapy as the primary approach for the analysis since it is a well-known approach to pain management. To make it possible to obtain the predictions, the values from the responses to the future interventions were one-hot encoded. Only psychotherapy was selected for the observation, resulting in the binary array where 1 is "ready to engage in psychotherapy", and 0 is "not ready to engage in psychotherapy", respectively.

### 4.4    **Data Analysis**

Our investigation delved into predicting readiness to engage in psychotherapy through the lens of linguistic characteristics, employing advanced computational analysis rooted in Natural Language Processing (NLP) and Machine Learning (AI). The Whisper framework, an open-source tool (accessible at https://github.com/openai/whisper), facilitated the acquisition of transcripts without external data sharing. Transcripts, sourced from participants responding to open-ended questions since June 18, 2023, were meticulously processed, combined with questionnaire responses, and standardized for analysis.

### 4.41  *Model Selection for Hypothesis Testing*

For hypothesis testing, we employed the GPT4All 13B snoozy, an innovative Large Language Model (LLM) based on LLama (https://opencompass.org.cn/model-detail/LLaMA-13B) and developed by NomicAI (https://www.nomic.ai/). Configured as a binary classifier through pre-prompting, the model labeled narratives as "not ready to engage in psychotherapy" (0) or "ready to engage in psychotherapy" (1). Predictions, showcasing high confidence, underwent both manual and algorithmic review.

### 4.42  *Hypothesis Testing Procedure*

Each hypothesis was tested using specific prompts, employing "one-shot learning" to elicit natural language predictions. To assess the performance of the binary classification model, we calculated sensitivity and specificity. Sensitivity measures the model's ability to correctly identify those ready to engage in psychotherapy, while specificity measures its accuracy in identifying those not ready to engage in psychotherapy. These metrics are fundamental for evaluating the model's predictive capacity and are essential additions to our comprehensive analysis. Manual validation translated these predictions into binary values (1 for "ready to engage in psychotherapy", 0 for the opposite). Despite the model's unexpected detailed annotations, manual review was viable due to the sample size. However, scalability concerns highlight the need for alternative approaches in larger datasets.

### 4.43  *Binary Classification Evaluation*

The manual validation results, encompassing the entire dataset of 24 observations, were converted into a binary array for critical metric evaluation. Using the pycm framework (https://www.pycm.io/) which facilitates hypothesis testing by generating confusion matrices and key metrics from binary predictions, we assessed essential metrics such as sensitivity and specificity, aligning with our classification approach. These metrics are fundamental for evaluating the model's predictive capacity in the binary classification task.

**4.44**    *Correlation Analysis between Text Features and Outcome Variable*

Our study explored correlations between text features and the outcome variable, readiness to engage in psychotherapy. Utilizing the Spacy framework (https://spacy.io/) for text feature analysis, we extracted keywords and created a correlation matrix, employing Matplotlib and Seaborn libraries. Top 30 keywords were correlated with variables such as Word Count, Sentence Count, and Average Word Length, offering insights illustrated in Table 2 and Table 3. These findings contribute to understanding the intricate relationships within our studied variables.

## 5. Results

### 5.1    Hypothesis A: Narrative-Based Perception of Pain Source

In testing Hypothesis A, which explored the narrative-based perception of pain source, participants were prompted to predict a person's readiness to engage in psychotherapy based on their chronic pain causation story. The accuracy (ACC) of the predictions achieved an impressive 95.65%, with a Specificity of 0.8 and a Sensitivity of 1. The Receiver Operating Characteristic (ROC) curve for binary classification further reinforced the robustness of the model, registering at 0.90 (see figure 1).

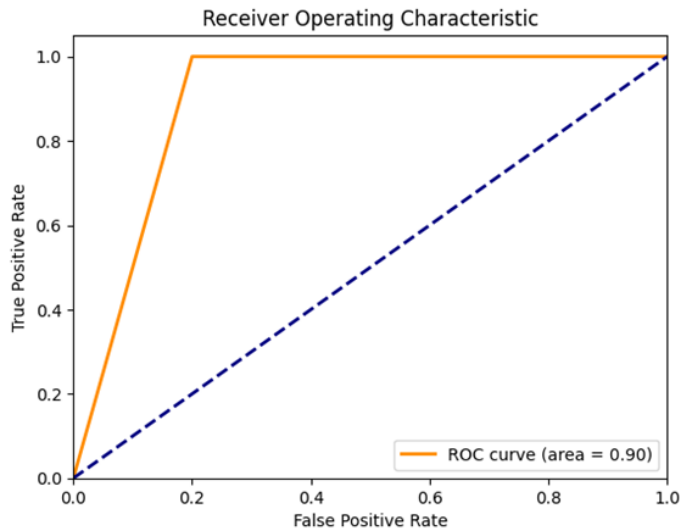

**Figure 1**: ROC Curve Hypothesis A: Narrative-Based Perception of Pain Source

### 5.1.1 *Correlation Matrix of Text Features and Psychotherapy in the Future*

In examining the relationship between "Psychotherapy_future" and "Sentence_count," a correlation coefficient of (r=0.54, p<0.01) reveals a moderate positive correlation. This means that an uptick in "Psychotherapy_future" is positively associated with a rise in "Sentence_count."

Similarly, the correlation between "Psychotherapy_future" and "Word_count" stands at (r=0.39, p<0.05), indicating a moderate positive correlation, though this association is somewhat less robust (see table 2).

| | Psychotherapy_future | Avg_word_length | Sentence_count | Word_count |
|---|---|---|---|---|
| Psychotherapy_future | — | | | |
| Avg_word_length | -0.042 | — | | |
| Sentence_count | **0.54 | -0.11 | — | |
| Word_count | *0.39 | -0.012 | 0.83 | — |

**Table 2:** Correlation table of text features of **Narrative-Based Perception of Pain Source** and psychotherapy in the future (N=23). **p<0.01, *p<0.05

## 5.2 Hypothesis B: Narrative-Based Pain Interference

In testing Hypothesis B, which investigated narrative-based pain interference, participants were tasked with predicting readiness to engage in psychotherapy based on the impact of chronic pain on an individual's life. Despite the conversion of annotated results to binary labels, the model displayed a macro accuracy (ACC) of 65.22%. Specificity stood at 0.2, while Sensitivity at 0.8. The corresponding ROC curve for binary classification was 0.49 (see figure 2).

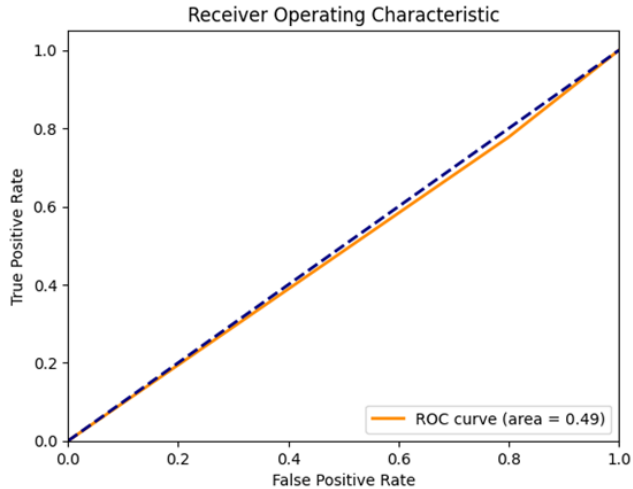

**Figure 2:** ROC Curve Hypothesis B: Narrative-Based Pain Interference (N=23).

## 5.3 Hypothesis C: Narrative-Based Factors Affecting Pain

In testing Hypothesis C, focusing on narrative-based factors affecting pain, participants predicted future readiness to engage in psychotherapy based on stories about chronic pain triggers and coping mechanisms. Results, again converted from annotated predictions to binary labels, showed an accuracy (ACC) of 83.33%. Specificity was 0.6, and Sensitivity resulted in 0.9. The ROC curve for binary classification reflected a substantial 0.75 (see figure 3).

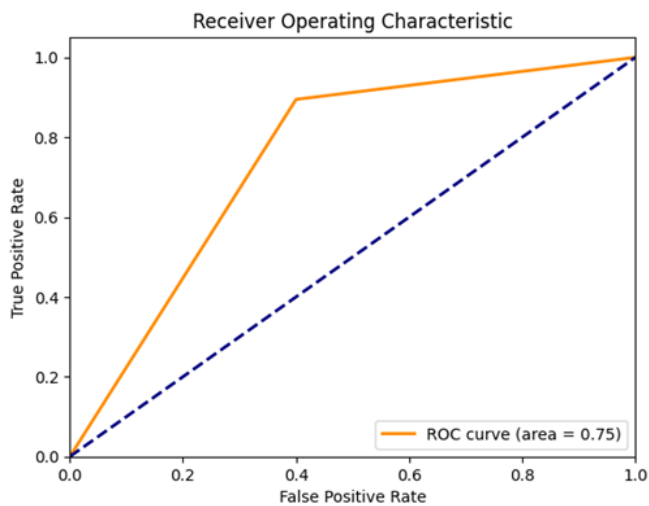

**Figure 3**: ROC Curve Hypothesis C: Narrative-Based Factors Affecting Pain (N=24).

### 5.3.1 *Correlation Matrix of text features and psychotherapy in the future*

In examining the relationship between "Psychotherapy_future" and "Sentence_count," a correlation coefficient of ($r=0.24$, $p<0.5$) reveals a weak positive correlation. This means that an uptick in "Psychotherapy_future" is moderately associated with a rise in "Sentence_count" (see table 3).

| | Psychotherapy_future | Avg_word_length | Sentence_count | Word_count |
|---|---|---|---|---|
| Psychotherapy_future | — | | | |
| Avg_word_length | -0.19 | — | | |
| Sentence_count | *0.24 | -0.27 | — | |
| Word_count | 0.14 | -0.33 | 0.92 | — |

**Table 3:** Correlation table of text features of Narrative-Based Factors Affecting Pain and psychotherapy in the future (N=24). *p<0.5

## 6. Discussion:

This study aimed to explore whether recorded pain narratives can predict one's readiness to engage in psychosocial interventions, specifically psychotherapy, by collecting and analyzing voice recordings of chronic pain patients, with NLP and Machine learning approaches. Findings demonstrate the model's competence in predicting readiness to engage in psychotherapy based on narrative content, providing valuable insights into the factors influencing individuals with chronic pain in their pain management. Notably, the model demonstrated high accuracy in using perception of pain source (95.65%) and factors affecting pain (83.33%) to predict readiness to engage in psychotherapy while pain interference demonstrated a low accuracy of 65.22%.

Differences in prediction outcome may be due to underlying differences in complexity and underlying relationships of these pain aspects and the model's ability to understand and make sense of the decisions or predictions made by the model. The robust accuracy in pain-related narratives of perception of pain source and factors affecting pain may be due to the connectedness to one's psychological state while considering engaging in psychotherapy (Scott et al., 2016). This may indicate that certain features related to perception of pain source and factors affecting pain, are more strongly indicative of psychological readiness, leading to higher predictive accuracy. In pain interference narratives prediction revealed low accuracy, possibly connected to the multifaceted nature of pain related interference influenced by various factors, including middle age, self-reported mental health, and different levels of pain interference at different stages of illness (Putzke et al., 2002), suggesting that pain interference is not a singular or isolated construct but is interconnected with demographic and mental health variables. The wide-ranging scope of pain interference might make it challenging to pinpoint specific aspects directly related to an individual's psychological readiness or willingness to engage in psychotherapeutic interventions. In sum, these results provide insights into the complexities of the pain experience and possible associations between the pain experience and readiness to engage in psychotherapy.

## 6.1    Integration into the Broader Context of Psychosocial Interventions for Chronic Pain

Placing our study within the broader landscape of psychosocial interventions for chronic pain reveals its alignment with the evolving field. The model's ability to predict engagement through narrative analysis resonates with the increasing emphasis on personalized, patient-centered healthcare. Existing research establishes the effectiveness of psychosocial interventions, encompassing various psychotherapeutic approaches, in managing chronic pain conditions (Driscoll et al., 2021). However, a patient's readiness to engage in these interventions relies on their readiness to first be aware and accept psychosocial explanations for their pain, find the intervention appealing and have access to it (Garrett et al., 2021). In addition, patient's beliefs about treatment were identified as main barriers to non-pharmacological pain treatment modalities (Becker et al., 2017). It is important to therefore understand the different factors of readiness. These factors include disinterest in the approach, willingness to work in therapy, willingness to discuss personal matters, and concern about one's problems. Taken together, these factors collectively indicate a person's psychological state of being ready for psychotherapy (Ogrodniczuk et al., 2009). Our study further underscores the necessity of nuanced assessments of pain perceptions to address the multifactorial nature of readiness for psychosocial interventions. The model's high accuracy suggests its effectiveness in capturing meaningful patterns related to pain source perception, the impact of pain on life, and factors influencing pain intensity. This indicates that considering a spectrum of factors, including triggers, coping mechanisms, and beliefs about pain source, contributes to a more holistic understanding of an individual's readiness for psychotherapy.

## 6.2    Assessing Perceptions and Readiness

Assessing one's perceptions provides valuable insight to one's readiness to accept biopsychosocial explanations for pain and readiness for psychotherapy and highlights the significance of beliefs in shaping treatment expectancies.  The process of change varies between and within individuals over time and is based on how one conceptualizes their condition (Caneiro et al., 2023). Recent research showed that individuals attributing their depression to biomedical causes tend to have pessimistic expectations about the success of treatment. This pessimism, in turn, is associated with negative expectancies about forming a strong therapeutic alliance with a therapist (Lebowitz et al., 2021). Moreover, holding a multifactorial belief about the cause of pain is strongly associated with increased engagement in pain self-management (Hardman et al., 2019), indicating the need for ongoing reinforcement of biopsychosocial explanations for pain cause, possibly by healthcare providers over time. In the context of clinician-patient communication, recent research shows that actively listening to a patient's subjective pain experience, specifically concerning uncertainty aspects of their pain, may help address the relational and psychological aspects of pain experience (Costa et al., 2023). In this sense, results reflect the importance of understanding patients' beliefs of the pain source for better assisting one's readiness to engage in psychotherapy.

## 6.3    The Role of Narratives and Metaphors

Narratives and metaphors emerge as valuable indicators shaping perceptions of pain for both patients and clinicians. They act as active components in the therapeutic process, enabling individuals to generate new meanings and cope with chronic pain (Chow & Fok, 2020). Previous research suggests that these metaphors, when clinically reinforced and embodied, become enactive, influencing the patient's experience (Stilwell et al., 2020). The focus on pain-related metaphors and their connection to clinical communication aligns with our exploration of linguistic markers in psychotherapy readiness. Recognizing enactive metaphors as a learning mechanism introduces a novel perspective, emphasizing narratives not merely as indicators but as active components shaping the therapeutic process and influencing tailored interventions.

## 6.4     Association Between Linguistic Features and Readiness for Psychotherapy

Our results explored correlations between linguistics features and readiness to engage in psychotherapy, revealing small to medium positive associations between the readiness for psychotherapy and number of the narrative sentences describing perceived causes of pain (Hyp A) and factors affecting pain (Hyp C). Our finding resonates with recent research underscoring the potential of Natural Language Processing (NLP) applications in mental health interventions seeking to understand the characteristics of patients, providers, and their relationships (Malgaroli et al., 2023), as well as understanding contextual factors and the ongoing need for innovative approaches in exploring processes of change in psychotherapy research (Gómez Penedo et al., 2023; Zilcha-Mano & Ramseyer, 2020). Furthermore, the evolution of context-sensitive analyses (Yin et al., 2019), particularly in response to the increased prevalence of digital platforms and the expansion of large corpora generated by telemedicine mental health interventions (MHI) (Fernandes et al., 2022), resonates with our study's exploration of the potential of NLP applications.

Our findings align with the progress observed in the current literature, showcasing advancements in the identification of contributors to treatment outcome based on one's level of readiness to change. Interestingly and contra intuitively, recent findings showed how chronic pain patients who scored lower on readiness to change, displayed more substantial improvements in pain outcomes from pre- to post-intervention in a CBT telehealth intervention compared to the exposed group with higher readiness to change, underscoring an intriguing prospect that those with initially lower readiness can benefit significantly from certain psychotherapeutic approaches (Romm et al., 2023). These findings emphasize the nuanced interplay of factors influencing therapeutic progress and treatment outcome and point to the complexity of the subjective pain experience and the importance of finding new and efficient research avenues to further explore pain perceptions and motivational aspects of chronic pain patients in their pain management. Our study's emphasis on understanding contextual factors and processes of change is consistent with the proposed integration of various contributions into a unified framework (NLPxMHI) for mental health service innovation and better personalized based treatment plans.

In our study, examining relationships between text features and readiness for psychotherapy highlighted valuable insights into the narrative characteristics associated with readiness for psychotherapy. This aligns with Fernández-Lansac and Crespo's findings, indicating that narratives about negative and traumatic events tend to be longer than those about neutral or positive experiences (Fernández-Lansac & Crespo, 2015). This insight is significant in understanding how individuals express both emotional and physical pain. Furthermore, it contributes to the evolving field of content analysis, enhancing the prediction of perceptions and deepening our understanding of pain subjectivity and its connection to readiness for change. The positive correlation implies that individuals displaying openness to psychotherapy are inclined to provide more detailed narratives regarding pain source and factors affecting pain, underscoring the potential significance of narrative richness as an indicator of psychological readiness for engaging in psychotherapeutic interventions.

## 6.5     The Fusion of NLP and Machine Learning in Psychotherapy Research

In summary, the fusion of natural language processing with psychotherapeutic insights presents a novel approach. The model shows promising potential in predicting readiness to engage in psychotherapy, highlighting the necessity for further refinement and validation (Zhou et al., 2022).

Merging natural language processing with psychotherapeutic insights offers a new approach and aims at predicting key psychotherapy processes, offering valuable insights into the challenges associated with linguistic processing, and providing practical recommendations for incorporating ML into psychotherapy research (Goldberg et al., 2020).

Recent study conducted by Guite et al. (2014) in the domain of chronic musculoskeletal pain among adolescents provides additional depth to our understanding of predictors and readiness for change. Delving into pain beliefs and readiness to change, their work aligns with the growing acknowledgment of narrative richness and elaboration as central factors in predicting psychological readiness for engagement in psychotherapeutic interventions. Examining pain-related beliefs and attitudes offers nuanced perspectives on how individuals articulate and perceive pain, contributing essential layers to the ongoing discourse on predicting treatment success and tailoring interventions effectively.

As we integrate these insights into the broader context of machine learning (ML) applications in psychotherapy research, a holistic understanding of narrative content, including pain-related beliefs, emerges as crucial. The recent study by Taubitz et al. (2022) underscores the transformative potential of ML in psychotherapy research, utilizing decision trees and ensembles to predict cognitive behavioral therapy outcomes. The research identified predictors such as baseline severity, somatization, and passive-aggressive traits, demonstrating that ML models can predict therapy outcomes and emphasizing the practical relevance of these predictors. The study aligns with the imperative for fine-tuning and personalization in psychotherapy, offering potential assistance to therapists in identifying complex cases that may require special attention.

## 6.6     Practical Significance and Caution in Generalization

These findings are practically significant for designing personalized interventions based on individual narrative profiles. This can improve the efficiency of psychotherapy and minimize resistance, supporting the wider objectives of patient-centered care. Yet, these results should be generalized cautiously, and more research is needed with larger, diverse populations.

## 7. Conclusion

In the exploration of the predictive capabilities of recorded pain narratives using Natural Language Processing (NLP) and Machine Learning (ML), this study has yielded significant insights into the readiness of individuals with chronic pain to engage in psychosocial interventions, specifically psychotherapy. The findings underscore the transformative potential of leveraging innovative technologies in understanding and enhancing patient-centric healthcare.

The ability to predict engagement and understand contextual factors of readiness to engage in psychotherapy through narrative analysis resonates with the broader shift toward personalized and patient-centered healthcare and aligns with the evolving landscape of psychosocial interventions for chronic pain.

The practical significance of these findings lies in the potential for tailored interventions based on individual narrative profiles. Understanding a patient's readiness for psychotherapy can enhance the effectiveness of interventions and predict an engagement and readiness to the future psychosocial interventions, such as psychotherapy, aligning with the broader goals of patient-centric care.

In conclusion, while the model demonstrated promising predictive capabilities in the context of psychotherapy engagement, the study underscores the need for ongoing refinement and validation. The fusion of natural language processing with psychotherapeutic insights presents a novel approach, but the practical application requires careful consideration of contextual nuances and ongoing advancements in the field.

**7.1     Limitations and Future Research**

It is essential to acknowledge certain limitations, such as the reliance on self-reported narratives and the inherent subjectivity in interpreting Large Language Model predictions. Additionally, the model's performance may vary based on individual differences in narrative expression and the complex nature of pain experiences. Caution is advised in generalizing these findings, given the need for more extensive research involving larger and diverse populations. The study presents a significant step forward but is a part of a larger journey toward a comprehensive understanding of readiness for psychosocial interventions.

For predictive modeling, the GPT4All 13B snoozy model, an extension of the LLama framework, was a strong choice for testing hypotheses. The model's 'one-shot learning' approach, which involved using specific prompts to guide the model in predicting psychotherapy engagement based on narrative content, showcased its flexibility and ability to handle varied inputs. However, it's important to note that the model's predictions were in natural language, requiring further manual validation and extraction of values. Besides that, since Large Language Models deeply rely on the 'prompt' construction process and trained data, there is more space for future research and additional study may be performed.

Despite the fact that obtained results are providing promising results, it is necessary to highlight that the current approach was also chosen due the fact of the small sample size, that made it impossible to use classic Machine Learning methods such as SVM or Logistic Regression. To resolve that issue, considering not only the small sample size but also imbalanced classes (the predictor variable), LLM was considered as a good starting point since its ability to leverage the context and subtext of the given texts. However, future exploration and experiments with different models may be performed, to ensure better outcomes and deliver reliable and reproducible results.

# 9. Appendix

## 9.1 Consent Form

PAINSTORY: NARRATIVES OF CHRONIC PAIN PATIENTS

Please read through this consent form and respond to the questions below it. Your consent needs to be given before moving on to the task. Estimated participation time 15-20 minutes.

We ask you to participate in our study "PainStory". Your participation may significantly contribute to deeper understanding of chronic pain.

STUDY PURPOSE: PainStory research collects pain narratives around the world using a new updated application to gain a deeper understanding of pain to improve diagnostic tools, relieving suffering and reducing stigma of chronic pain.

Why am I being invited to take part in a research study?

We are inviting you to take part in this study because you are 18+ years old, you have a chronic pain condition and you are interested in participating.

What should I know about participation in a research study?

- Whether or not you take part is up to you
- You can choose not to take part
- You can agree to take part and later change your mind
- Your decision will not be held against you

What happens if I say yes, I want to be in this research study?

You will be asked to:

Specify your pain condition, rate your pain level, complete questionnaires and share your pain experience while our web system records your narrative.

How long will the research last?

We expect the completion of the study to take approximately 15-20 minutes.

What happens if I say yes, but I change my mind later?

Participation in this study is completely voluntary. You can **leave the research at any time** it will not be held against you. Data collected to the point of withdrawal may still be used for data analysis. However, if you don't want to share with us your pain narrative after recording it, you may start over the recording, replacing your initial recording with an empty one. Your data is always kept de-identified and confidential.

Is there any way being in this study could be bad for me?

There is no potential harm to you.

What happens to the information collected for the research?

Data will be kept on AWS and Azure cloud services. Recordings will be kept separate from self report responses. We will de-identify clinical data extracted from personal data and Protected Health Information (PHI, which is individually identifiable health information), and store data collected in a manner aligned with the University Ethics Committee (IRB, which is the Institutional Review Board) ethics requirements and the laws in which we are obliged to. Organizations that may inspect and copy your information include the IRB and other representatives of this organization.

We may retain your data after the study for future research. It will be stored on an encrypted server on a designated cloud server storage that is used exclusively for this study. Data will be de-identified so that it contains no information indicating your name or personal identifiable information.

Will being in this study help me in any way?

Benefits of participation include any satisfaction you may feel from contributing to scientific research, also sharing your pain narrative might have a therapeutic effect.

Who can I talk to?

If you have any questions, concerns, or complaints; or you feel the research has hurt you, you can contact the research team at the Integrative Pain Lab at the University of Haifa, School of Public Health. The lab can be reached at ipain@research.haifa.ac.il

This research has been reviewed and approved by an Institutional Review Board ("IRB"). You

may talk to them at 04-824-9948 or via ethics-health@univ.haifa.ac.il **if**:

- Your questions, concerns, or complaints are not being answered by the research team
- You cannot reach the research team
- You want to talk to someone besides the research team
- You have questions about your rights as a research subject
- You want to get information or provide input about this research

RECORDING(S) RELEASE CONSENT FORM

Your recording(s) will not be posted on any public website and cannot be downloaded by any computer user on or off campus without your permission. Also, your name and any other personally identifying information will not be associated with your recording(s).

Please indicate below your permissions for the Integrative Pain Lab at the University of Haifa (iPainLab) to use your recording(s):

- o I'm 18+ years of age [REQUIRED]

- o I grant permission to use my audio recordings (anonymously) for research purposes [REQUIRED].

- o I grant permission to use recording(s) of me in scientific presentations (anonymously) .

- o I speak English fluently

DO YOU CONSENT TO PARTICIPATE IN THIS RESEARCH STUDY?

- o YES, I AGREE TO PARTICIPATE

- o NO, NOT INTERESTED